\documentclass[aip,jcp,reprint]{revtex4-2}
\usepackage{amsmath, amssymb}
\usepackage{graphicx}
\usepackage{hyperref}
\usepackage{listings}
\usepackage[frozencache,cachedir=.]{minted} % [frozencache,cachedir=.]
\usepackage{xcolor}
\usepackage{physics}
\usepackage{caption}
\usepackage{subcaption}
\usepackage{orcidlink}
\usepackage[nolist,nohyperlinks]{acronym}

\hypersetup{
    colorlinks=true,
    linkcolor=blue,
    citecolor=blue,
    urlcolor=blue
}

\begin{document}

\title{QMCkl: A Kernel Library for Quantum Monte Carlo Applications}

\newcommand{\LCPQ}{Laboratoire de Chimie et Physique Quantiques (UMR 5626), Universit\'e de Toulouse, CNRS, France}
\newcommand{\LCPT}{Universit\'e de Lorraine, CNRS, LPCT, Nancy, France}
\newcommand{\MESA}{MESA+ Institute for Nanotechnology, University of Twente, P.O. Box 217, 7500 AE Enschede, The Netherlands}
\newcommand{\AMU}{Aix Marseille Univ, CNRS, Centrale Méditerranée, iSm2, Marseille 13013, France}
\newcommand{\UVSQ}{Universit\'e Paris-Saclay, UVSQ, LI-PaRAD, France}
\newcommand{\LCT}{Sorbonne Universit\'e, LCT, UMR 7616 CNRS, 75005 Paris, France}
\newcommand{\CINECA}{CINECA, Via Magnanelli 6/3, Bologna, 40033, Italy}

\author{Emiel Slootman\orcidlink{0000-0003-3228-4055}}  %%
\affiliation{\MESA{}}

\author{ Vijay Gopal Chilkuri\orcidlink{0000-0003-4827-3588}}  %%
\affiliation{\LCPQ{}}
\affiliation{\AMU{}}

\author{ Aurelien Delval\orcidlink{0000-0002-2707-6940} } %%
\affiliation{\UVSQ{}}

\author{ Max Hoffer } %%
\affiliation{\UVSQ{}}

\author{ Tommaso Gorni } %%
\affiliation{\CINECA{}}

\author{ François Coppens\orcidlink{0000-0003-1695-352X} } %%
\affiliation{\LCPQ{}}
\affiliation{\UVSQ{}}

\author{ Joris van de Nes} %%
\affiliation{\MESA{}}

\author{ Ramón L. Panadés-Barrueta\orcidlink{0000-0003-4239-0978} }  %%
\affiliation{\MESA{}}

\author{ Evgeny Posenitskiy\orcidlink{0000-0002-1623-0594} } %%
\affiliation{\LCPQ{}}

\author{ Abdallah Ammar\orcidlink{0009-0009-2774-0520} } %%
\affiliation{\LCPQ{}}
\affiliation{\LCPT{}}

\author{Edgar Josué Landinez Borda\orcidlink{0000-0003-3903-4809}} %%
\affiliation{\MESA{}}

\author{ Kevin Camus} %%
\affiliation{\UVSQ{}}

\author{ Oto Kohulàk\orcidlink{0000-0002-5933-5804} } %%
\affiliation{\LCPQ{}}
\affiliation{Computing Center, Centre of Operations of the Slovak Academy of Sciences, 84535 Bratislava, Slovakia}

\author{ Emmanuel Giner\orcidlink{0000-0002-6206-1103}} %%
\affiliation{\LCT{}}

\author{ Pablo de Oliveira Castro\orcidlink{0000-0001-9007-6145} } %%
\affiliation{\UVSQ{}}

\author{ Cedric Valensi\orcidlink{0000-0001-7187-5638} } %%
\affiliation{\UVSQ{}}

\author{ William Jalby\orcidlink{0000-0002-4975-5469} } %%
\affiliation{\UVSQ{}}

\author{Claudia Filippi\orcidlink{0000-0002-2425-6735}}  %%
\affiliation{\MESA{}}

\author{Anthony Scemama\orcidlink{0000-0003-4955-7136} }  %%
\email{scemama@irsamc.ups-tlse.fr}
\affiliation{\LCPQ{}}

\begin{abstract}
Quantum Monte Carlo (QMC) methods deliver highly accurate electronic structure calculations but are computationally intensive. The quantum Monte Carlo kernel library (QMCkl) provides a modular, portable collection of high-performance kernels implementing the core building blocks of QMC calculations. It offers a C-compatible API, supports the TREXIO standard for input, and covers essential QMC kernels including atomic and molecular orbitals, cusp corrections, Jastrow factor, and the necessary derivatives also to  perform variational and structural optimization. QMCkl separates algorithmic development from hardware-specific tuning by combining human-readable reference implementations with performance-optimized kernels that produce identical numerical results. The library enables consistent, efficient, and reproducible simulations across different QMC codes and architectures, and achieves substantial speedups in the evaluation of the energy and its derivatives. Beyond QMC, QMCkl can accelerate deterministic quantum chemistry workflows and visualization tools, promoting cross-code interoperability and simplifying high-performance scientific software development.
\end{abstract}

\maketitle

\section{Introduction}
Quantum Monte Carlo (QMC) methods\cite{QMC1, QMC2, QMC3} provide a stochastic solution to the many-body Schr\"odinger  equation and are widely recognized for their high accuracy in predicting the electronic properties of molecular and extended systems. Even though QMC methods are computationally demanding, they are naturally suited to parallelization, making them ideal candidates for high-performance computing (HPC) platforms. Nevertheless, their complexity often results in monolithic implementations with limited modularity, portability, or ease of maintenance.

The quantum Monte Carlo kernel library\cite{qmckl_github} (QMCkl) developed within the TREX (Targeting REal chemical accuracy at the eXascale) Center of Excellence in Exascale Computing\cite{trex}, aims to adapt and scale real-space QMC codes to leverage the full potential of
%next-generation
high-performance computing systems. Instead of optimizing each code separately, QMCkl provides a unified, portable, and efficient foundation, ensuring faithful numerical behavior across codes as demonstrated here for CHAMP\cite{CHAMP-EU} and QMC=Chem\cite{qmcchem,scemama_2013}. Its modular design also enables deterministic quantum chemistry programs, such as Quantum Package\cite{garniron_2019}, to reuse the same kernels for evaluating Jastrow factors in transcorrelated calculations, or molecular orbitals for the computation of density grids.

The library’s design follows a principle of separation of concerns, inspired by classical high-performance numerical libraries such as BLAS\cite{blas3} and LAPACK\cite{lapack}. Scientists write kernels in a clear, human-readable \emph{pedagogical version} in Fortran focused on correctness and readability, while HPC specialists implement the computationally intensive kernels in a C high-performance \emph{HPC version} optimized for speed. Both versions produce identical numerical results and can be accessed through the same C \ac{API}. The modular structure promotes clarity, maintainability, and reproducibility, allowing all QMC codes to benefit from the same verified and optimized routines.
The importance of reusable, interoperable scientific libraries has recently been highlighted in a broader context by Lehtola, who argues that such shared infrastructures are essential for reproducibility, sustainability, and the rapid dissemination of new computational methods.\cite{lehtola_2023}

The benefits extend beyond efficiency and reproducibility. By decoupling algorithmic development from low-level performance optimization, QMCkl allows scientists to retain full control over their scientific workflows while using their preferred programming language. Its C-compatible \ac{API} ensures interoperability with languages such as Python, Fortran, and C\texttt{++}, facilitating broad adoption. This design ensures that the scientific code remains transparent and accessible, while performance-critical components are handled within the library. 

The design allows QMCkl to adapt to new hardware without requiring complete rewrites of existing codes. When performance issues arise on new architectures, they can be addressed by updating the library itself, preserving the original scientific implementations. This allows QMC researchers to focus on their models and algorithms, while HPC specialists handle performance optimization. As a result, development is faster, codes remain reliable across hardware generations, and research teams can achieve high performance without deep HPC expertise.

In this work, we present the structure and design principles of QMCkl, describe its key computational kernels, demonstrate its performance and integration with existing QMC codes, and present several usage examples illustrating its interface and capabilities also for deterministic quantum chemistry and visualization applications. 
% We conclude by discussing ongoing developments and future plans for the library.

\section{Library Structure}\label{sec:library_structure}

%This section describes the main computational kernels implemented in QMCkl. The library provides the essential building blocks for evaluating Jastrow–Slater wave functions (Eq.~\eqref{eq:wf}) in both variational (VMC) and diffusion Monte Carlo (DMC) simulations. It enables the computation of the quantities required for all-electron and single-electron moves in the Monte Carlo walk. In addition, it includes derivatives with respect to Jastrow parameters for wave function optimization, and gradients with respect to atomic positions for the evaluation of interatomic forces.

A distinctive feature of quantum Monte Carlo methods is the repeated evaluation of several computational kernels at every Monte Carlo step. Typical kernels include the evaluation of \acp{AO}, \acp{MO}, Jastrow correlation factors, \textit{etc}.
Each of these kernels is written by a QMC specialist in a human-readable form in Fortran, referred to as the \emph{pedagogical version}. This version serves as a clear, interpretable reference implementation, much like the original Netlib implementations of BLAS, emphasizing correctness and readability over computational efficiency.
The choice of the Fortran language is motivated by its mathematical clarity, widespread familiarity in scientific computing, and natural syntax for array and linear algebra operations, making it an ideal language for documenting the intended algorithmic behavior. While these implementations are not optimized for performance, they provide a trusted baseline for verification and validation.
In contrast, the most computationally intensive kernels are re-implemented by HPC experts in an \emph{HPC version} in the C language, which prioritizes speed and efficiency through low-level optimizations, vectorization, and memory layout tuning. The HPC version is not expected to be readable or maintainable in the same way as the pedagogical version; its primary objective is to deliver high performance while producing identical numerical results, ensuring that the optimized code remains functionally equivalent to the reference. 
To ensure internal interoperability, both QMCkl versions can be accessed through the same C standard \ac{API}.

In the following, we describe the common development principles to both implementations and their structures.

\subsection{Literate Programming}
QMCkl is developed using the principles of literate programming\cite{knuth_1992}, where source code and documentation are embedded in the same files. In this approach, algorithms are first described in natural language and equations before being expressed as compilable code. This ensures consistency between documentation and implementation while improving readability. The library relies on \texttt{org-mode} \cite{schulte_2012,orgmode}, which provides a framework to generate human-readable documents (e.g. Portable Document Format (PDF) or Hyper-Text Markup Language (HTML)) and compilable source code from the same file.
%The source code can be conveniently extracted using Emacs\cite{emacs}, invoked via command-line within the \texttt{Makefile}, and subsequently compiled. Additionally, Emacs enables interactive execution of code blocks, akin to Jupyter notebooks\cite{Kluyver_2016}, but with the added benefit of multi-language support.
Because the Org sources are plain text, they integrate cleanly with version control systems and allow meaningful diffs and merges, unlike JSON-based notebook formats. Org’s explicit support for tangling (code extraction) and weaving (document generation) ensures that both the narrative and the corresponding source code remain synchronized and transparently linked.
Both the pedagogical and HPC implementation are combined in the Org source files, ensuring consistency between the reference and optimized versions. This methodology facilitates validation of the algorithms and maintains a transparent link between theory and implementation. 
While \texttt{org-mode} is widely used in research notebooks and reproducible reports, its application as the authoritative source for a performance-critical HPC scientific library is still uncommon. It is worth mentioning that QMCkl represents one of the first large-scale deployments of a fully literate-programming workflow in quantum chemistry, combining human-readable algorithms with architecture-optimized C kernels generated from the same Org source.

\subsection{Context and Call Structure}
%Each kernel relies on a set of intermediate quantities, many of which are shared across different kernels. For instance, electron–nucleus distances are needed both in the evaluation of orbitals and in the Jastrow factor.
%The design of QMCkl ensures that such intermediates are never recomputed unnecessarily. To achieve this, every Monte Carlo step is associated with a time stamp that uniquely identifies the state of the system. When an intermediate quantity is requested for the first time during a given step, QMCkl computes it and stores it in memory. If the same quantity is requested again within the same step, the precomputed value is retrieved directly. This mechanism is analogous to the dependency management of the \texttt{make} utility: only intermediates that depend on updated data are recomputed, while all others are reused.
%Whenever the electron coordinates are modified, the time stamp is incremented to indicate that all dependent intermediates may require updating.
%This dependency-driven, lazy evaluation strategy ensures that computations are performed only when necessary, avoiding redundant work and improving efficiency.
%
% Changes by Pablo
Each kernel relies on a set of intermediate quantities, many of which are shared across different kernels. For instance, electron–nucleus distances are needed both in the evaluation of orbitals and in the Jastrow factor. The design of QMCkl ensures that such intermediates are never recomputed unnecessarily. To achieve this, every Monte Carlo step is associated with a time stamp that uniquely identifies the state of the system. When an intermediate quantity is requested for the first time during a given step, QMCkl computes it and stores it in memory. If the same quantity is requested again within the same step, the precomputed value is retrieved directly. This approach implements a form of memoization,\cite{michie_1968} where function results are cached and reused: only intermediates that depend on updated data are recomputed, while all others are reused. Whenever the electron coordinates are modified, the time stamp is incremented to indicate that all dependent intermediates may require updating. This dependency-driven, lazy evaluation strategy ensures that computations are performed only when necessary, avoiding redundant work and improving efficiency.

\begin{figure}
    \centering
    \includegraphics[width=0.5\linewidth]{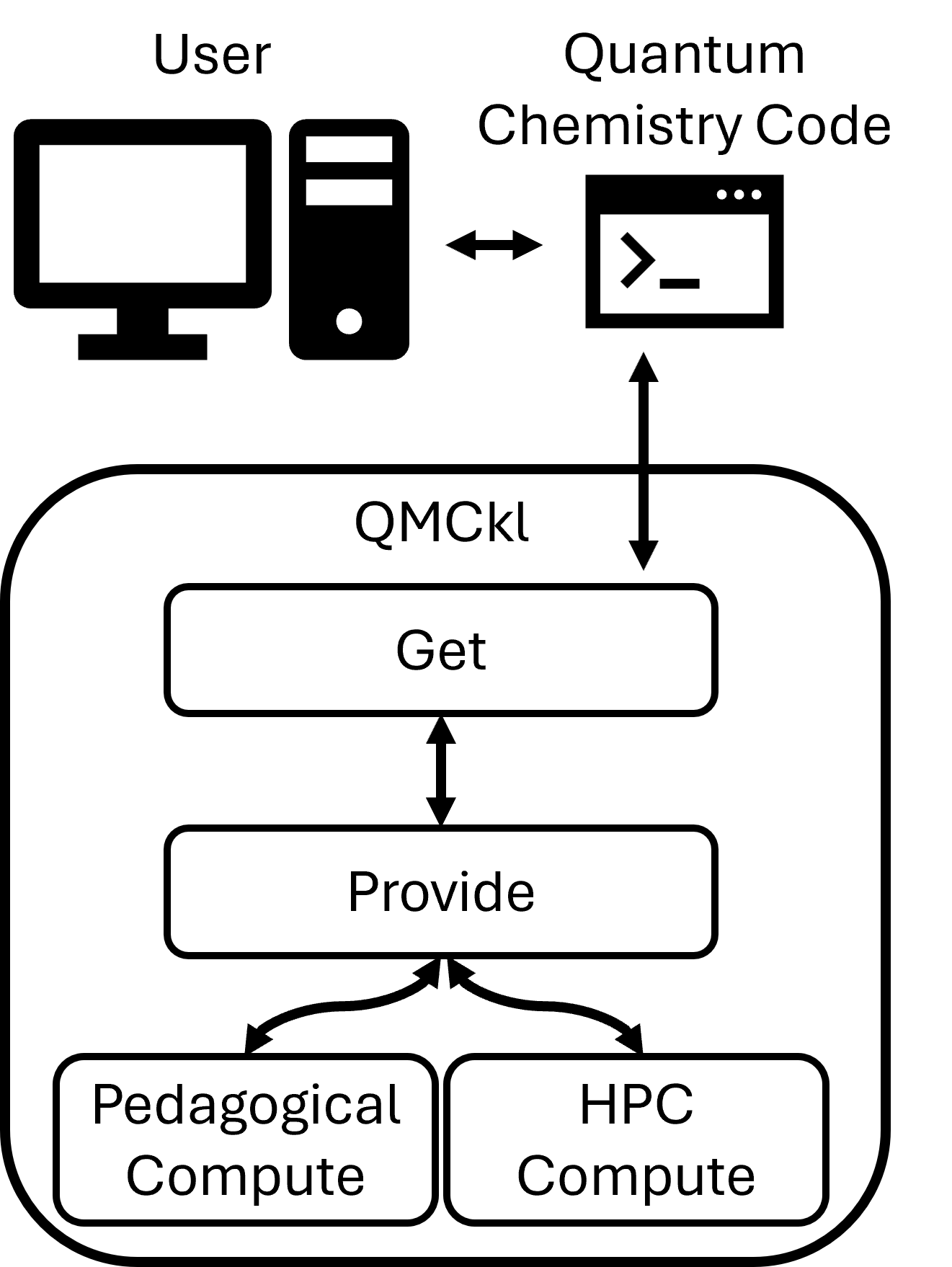}
    \caption{Call structure of the QMCkl library.}
    \label{fig:diagram}
\end{figure}

This call structure is schematically depicted in Fig.~\ref{fig:diagram}. A quantum chemistry code calls the \mintinline{shell}{get} function. This function in turn invokes the  \mintinline{shell}{provide} function, whose role is to guarantee that the requested quantity is valid within the current time stamp.
The \mintinline{shell}{provide} mechanism operates recursively. If the quantity is already valid, the function returns immediately. Otherwise, it first ensures the validity of all dependencies by calling their respective \mintinline{shell}{provide} functions, and then it invokes the corresponding \mintinline{shell}{compute} function, which evaluates the target quantity using the provided dependencies.~\cite{irpf90}
Depending on the compilation flags, the system can seamlessly switch between the pedagogical version or the HPC-optimized version of the compute functions.

All data related to target quantities and their intermediates are stored within a dedicated structure called the \emph{context}. A context represents the current state of the library, including all cached values and their associated time stamps. The library supports the creation of multiple independent contexts, allowing different calculations to be carried out simultaneously within the same program without interference.

\subsection{Cross-Language Support}
To ensure broad interoperability and maximize accessibility across diverse software ecosystems, core library functions and interfaces
%such as \mintinline{shell}{get} and \mintinline{shell}{provide}
are implemented primarily in the C programming language, while all internal functions corresponding to \mintinline{shell}{compute} also come in the pedagogical Fortran version.
The choice of C provides a well-defined, stable, and widely supported Application Binary Interface (ABI) that serves as a \textit{de facto} standard for cross-language function calling. By adhering to the C ABI, the library can be seamlessly integrated into systems written in a variety of languages, including C\texttt{++}, Fortran, Python, Rust, Julia, and others, via their respective Foreign Function Interface (FFI) mechanisms. This is made possible by the absence of name mangling, predictable calling conventions, and the availability of mature tooling for binding C interfaces. Furthermore, the use of \texttt{extern "C"} ensures ABI compatibility even when interfacing with C\texttt{++} code, while the reliance on simple, portable data types minimizes platform-specific complexities. As a result, writing the library in C and exposing an API with simple data types enables maximum portability, performance, and compatibility, making it a pragmatic choice for deployment in heterogeneous environments and facilitating adoption.

\subsection{Parallelization strategy}

QMCkl provides support for shared-memory parallelism through OpenMP.\cite{OpenMP6} 
The efficiency of this approach increases with the size of the problem, in particular when the number of electrons or the number of walkers is large. 
In such cases, arrays can be allocated with significant dimensions, which is not only beneficial for OpenMP efficiency but also prepares the kernels for future acceleration on graphics processing units (GPUs), where large data sets are required to fully exploit the available throughput.
As QMCkl maintains state through the context variable, it is important to note that a given context must not be accessed concurrently by multiple threads. Thread safety is ensured as long as each OpenMP thread operates on a distinct context.

Beyond shared-memory parallelism, QMCkl is designed to interoperate seamlessly with distributed-memory parallelism managed by the calling program via MPI. 
This allows users to adopt a hybrid MPI/OpenMP scheme in which a fixed number of cores is assigned to each MPI process, and OpenMP threads are employed within the process to exploit node-level or socket-level parallelism. 
Such flexibility makes it possible to tune the balance between inter-process and intra-process parallelism, and to adapt to the architectural characteristics of different high-performance computing platforms. 
This design ensures that QMCkl can be efficiently deployed on a wide variety of systems, from multicore CPUs to large-scale heterogeneous architectures targeted in exascale computing.

In addition to CPU optimizations, the library is being extended with GPU support. 
Dedicated kernels are under development with the objective of achieving portability across different GPU vendors. 
The modular structure of QMCkl, and its reliance on large contiguous arrays, naturally facilitates GPU porting.
These developments are part of the long-term roadmap to make QMCkl a core component of exascale-ready QMC software.

\subsection{TREXIO as standard input}
Although wave function parameters can be passed to QMCkl through multiple individual function calls, the library supports a streamlined and highly efficient alternative: all parameters can be loaded in a single call when the wave function is stored in the TREXIO\cite{trexio} format. This standardized file format, developed as part of the TREX project, serves as a unified input interface for key components of the trial wave function, including molecular geometry, basis set information, molecular orbital coefficients, and Jastrow parameters.  

TREXIO provides a unified C API with bindings available in many different programming languages, and supports both plain-text and HDF5 back-ends. This flexibility allows users to choose between human-readable formats, suitable for testing and debugging, and binary formats optimized for efficiency and scalability. By adopting TREXIO, QMCkl ensures seamless integration with codes in the QMC ecosystem, such as CHAMP\cite{CHAMP-EU}, QMC=Chem\cite{scemama_2013, qmcchem}, and TurboRVB\cite{nakano_2020}, as well as with deterministic quantum chemistry codes, including Quantum Package\cite{garniron_2019}, GammCor\cite{gammcor}, Dirac\cite{dirac}, Molpro\cite{molpro}, FHI-aims\cite{fhi-aims}, PySCF\cite{pyscf} and CP2K\cite{cp2k}. Furthermore, TREXIO also offers scripts to convert output of popular packages, notably GAMESS(US)\cite{gamess} and Gaussian\cite{gaussian}, to the TREXIO format.

\subsection{Build and Packaging}
QMCkl minimizes external dependencies to simplify installation and portability.
The library relies on standard numerical libraries (BLAS, LAPACK), with optional integration of TREXIO for input.
The build system is based on Autotools\cite{autotools}, chosen for its portability across high-performance computing platforms.
In addition to conventional source-based installation, QMCkl can be deployed through the GNU Guix package manager,\cite{guix} which enables fully reproducible builds and facilitates integration in controlled HPC environments.
Pre-compiled packages for common Linux distributions and Spack\cite{spack} support are planned to further facilitate adoption.
% As an integral component of the TREX software ecosystem, QMCkl integrates with complementary tools like TREXIO, a library that defines a standard format for wave function parameters and supplies input/output infrastructure across programming languages.\cite{trexio} Together, these libraries form a coherent framework enabling interoperable QMC implementations with an eye to future exascale performance.
%For diffusion Monte Carlo (DMC), the library implements the ingredients needed to handle nonlocal ECPs through the T-move algorithm.

\section{Computational Kernels}
\subsection{Atomic Orbitals}

Since the evaluation of \acp{AO} is a key step in QMC calculations, the computation is organized to maximize the reuse of intermediate quantities.
\acp{AO} are expressed as products of radial functions $R(\mathbf{r})$ and angular components $P(\mathbf{r})$, which are precomputed independently:
\begin{equation}
\chi_n (\mathbf{r}) = \mathcal{M}_n\, R_{\theta(n)} (\mathbf{r})\, P_{\eta(n)}(\mathbf{r}).
\end{equation}
$\theta(n)$ and $\eta(n)$ are integers that label, respectively, the shell in which the $n$-th \ac{AO} is expanded and its angular component, both taken from unique sets.
The factor $\mathcal{M}_n$ accounts for the possibility of using different normalizations within the same shell, as implemented, for instance, in GAMESS(US).
In addition to the orbital value $\chi_n(\mathbf{r})$, the library can also provide its gradient and Laplacian, $\nabla \chi_n(\mathbf{r})$ and $\nabla^2 \chi_n(\mathbf{r})$, with respect to the electron coordinates. 
%The kernels are designed to operate both on electron positions and on arbitrary grid points, making them suitable for visualization and DFT-style integration.

The radial part of a shell, $R_s(\mathbf{r})$, is expressed as a linear combination of $N_{\text{prim}}$ \emph{primitive} functions which can be of either Slater ($p=1$) or Gaussian ($p=2$) type, centered at position $\mathbf{R}_A$:
\begin{equation}
  R_s(\mathbf{r}) = \mathcal{N}_s \|\mathbf{r}-\mathbf{R}_A\|^{n_s}
  \sum_{k=1}^{N_{\text{prim}}} a_{ks}\, f_{ks}
 \exp \left( - \gamma_{ks} \| \mathbf{r}-\mathbf{R}_A \| ^p \right).
\end{equation}
This form is expected to have sufficient flexibility for using QMCkl with trial wave functions coming from different software.
The normalization factor $\mathcal{N}_s$ may be used to ensure that each radial function integrates to unity. Basis sets are usually specified as combinations of normalized primitives, which requires that the normalization coefficients of the primitives $f_{ks}$ are provided. The remaining coefficients, $n_s$, $a_{ks}$ and $\gamma_{ks}$, are supplied as part of the basis set.

The primary computational bottleneck in evaluating the radial functions arises from the exponential function. 
In QMCkl, a \emph{precision} parameter is stored within the context, specifying the number of bits of precision requested by the user. 
By default, this parameter is set to $B = 53$, corresponding to IEEE-754 standard double precision. 
Exponentials are computed only if their argument satisfies
\begin{equation}
    -\gamma_{ks} \, \| \mathbf{r}-\mathbf{R}_A \|^p > -\log(2^{B-1}) ,
\end{equation}
otherwise, the primitive and all its derivatives are assigned a value of zero.

For efficiency, all primitives are initially reordered so that those sharing the same center are sorted in ascending order of exponent. 
As a result, once the argument of an exponential is identified as negligible, all subsequent primitives at that center can be skipped.

%When $B < 24$, the exponentials are computed in single precision, providing an additional speedup.
To further accelerate the evaluation, each nucleus is assigned a \emph{radius}, defined as the maximal distance beyond which the primitive with the smallest exponent effectively vanishes. 
If $\| \mathbf{r}-\mathbf{R}_A \|$ exceeds this radius, all primitives associated with nucleus $A$ are skipped.
As the number of electrons in the vicinity of a nucleus is independent of the size of the molecule, the cost of the evaluation of the radial functions grows linearly with the number of atoms.

The angular function $P_l$ is defined as
\begin{equation}
    P_l(\mathbf{r}) = (x - X_A)^a (y - Y_A)^b (z - Z_A)^c ,
\end{equation}
with $\mathbf{r} = (x,y,z)$, $\mathbf{R}_A = (X_A, Y_A, Z_A)$, and angular momentum $\ell = a + b + c$. 
The triplet $(a,b,c)$ associated with a given basis function is determined by the mapping $\eta(n)$.

The evaluation of the angular part of the atomic orbitals can be performed very efficiently. 
First, the coordinate differences $(x - X_A)$, $(y - Y_A)$, and $(z - Z_A)$ are computed. 
For each Cartesian direction, powers of these differences up to order $\ell$ are then generated iteratively: starting from the zeroth power, successive values are obtained by multiplication with the corresponding coordinate difference. 
The resulting powers are stored in intermediate arrays within the context, and fetched when needed for the evaluation the \acp{AO}.

This iterative construction offers several advantages. 
It avoids repeated calls to the costly power function (\texttt{pow})  and, at the same time, provides all the powers required to assemble the polynomial combinations corresponding to a given angular momentum. 
Moreover, the same intermediates are reused in the computation of derivatives, which are needed for the gradients and Laplacians of the atomic orbitals.

\begin{figure}
    \centering
    \includegraphics[width=0.9\linewidth]{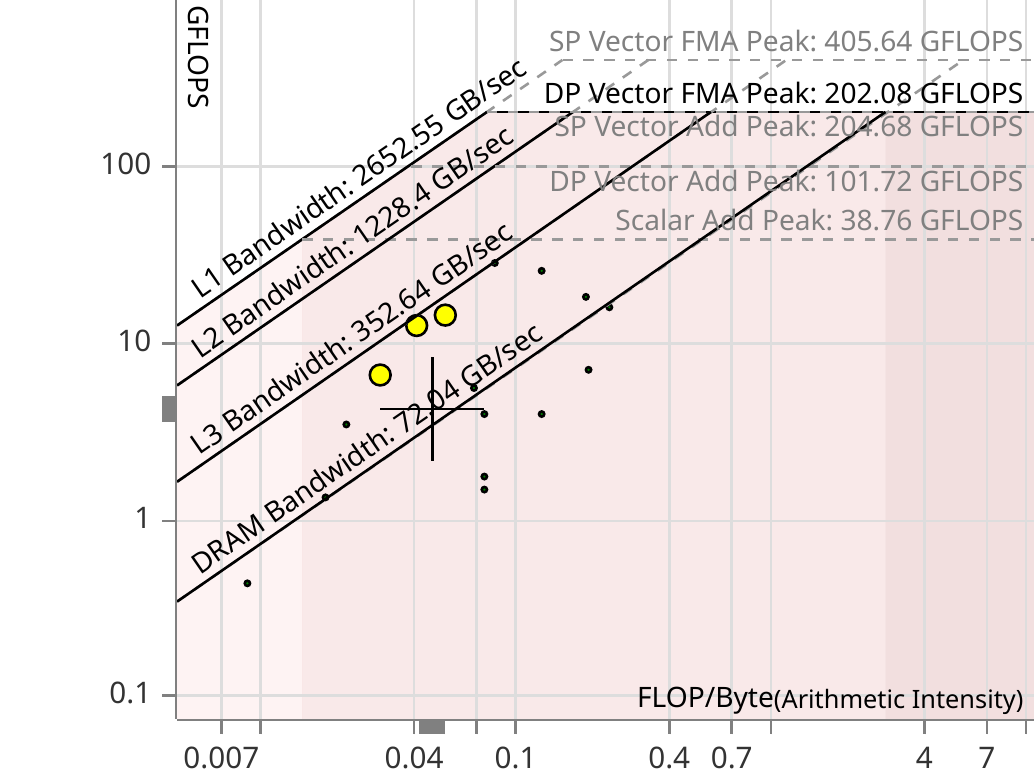}
    \caption{Roofline performance model of the atomic-orbital computation kernels. 
    The yellow dots represent the most time-consuming loops, while the cross indicates the average performance. Small dots correspond to less time-consuming loops.}

    \label{fig:ao_roofline}
\end{figure}

The performance of the kernels involved in the computation of atomic orbitals was evaluated on an Intel Core i7-13850HX processor.
The benchmark system is a Cu(II)–peptide complex ([C$_{34}$H$_{48}$CuN$_{11}$O$_{10}$]$^{2+}$), described with the cc-pVDZ basis set. This results in 1114 atomic orbitals constructed from 2650 primitive Gaussian functions. Figure~\ref{fig:ao_roofline} compares the measured performance with the theoretical hardware bandwidth and compute ceilings. The three dominant computational loops fall between the L3 and DRAM bandwidth rooflines, confirming that their execution is fully memory-bound.

\subsection{Molecular orbital kernels}

Building on the \ac{AO} routines, QMCkl provides kernels for the evaluation of \acp{MO}. 
The computation of \ac{MO} values, together with their gradients and Laplacians, constitutes one of the central kernels in QMC. 
Each MO, indexed by $n$, is expressed as a linear combination of \acp{AO} with coefficients $C_{nk}$:
\begin{align}
  \phi_n(\mathbf{r}) &= \sum_{k} C_{nk}\, \chi_k(\mathbf{r}), \\
  \nabla \phi_n(\mathbf{r}) &= \sum_{k} C_{nk}\, \nabla \chi_k(\mathbf{r}), \\
  \Delta \phi_n(\mathbf{r}) &= \sum_{k} C_{nk}\, \Delta \chi_k(\mathbf{r}),
  \label{eq:mos}
\end{align}
which naturally maps to matrix–matrix multiplications. 
Although dense matrix multiplication is among the most optimized operations in high-performance computing, the computational scaling of \ac{MO} evaluation is 
$\mathcal{O}(N_{\text{AO}} \times N_{\text{elec}} \times N_{\text{MO}})$,
i.e.\ the product of the numbers of \acp{AO}, electrons, and \acp{MO}. 
This scaling exceeds that of \ac{AO} evaluation, making careful optimization of this step essential.

A first optimization in QMCkl is to concatenate \ac{AO} values, gradients, and Laplacians into a single array, enabling their evaluation through a unique \ac{GEMM} call. 
This reduces overhead and maximizes the reuse of the $C$ matrix, but does not change the asymptotic scaling. 
A more substantial improvement was introduced in the QMC=Chem software~\cite{scemama_2013}, where the sparsity of AOs and their derivatives was exploited to reduce the scaling to 
$\mathcal{O}(N_{\text{elec}} \times N_{\text{MO}})$,  
while maintaining data structures optimized for vectorization and high arithmetic intensity. 
This algorithm has been integrated into the high-performance version of QMCkl. 
In addition, QMCkl enables the selection of subsets of \acp{MO}, allowing users to exclude unoccupied orbitals, which further reduces the effective $N_{\text{MO}}$.

\begin{figure}
    \centering
    \includegraphics[width=0.9\linewidth]{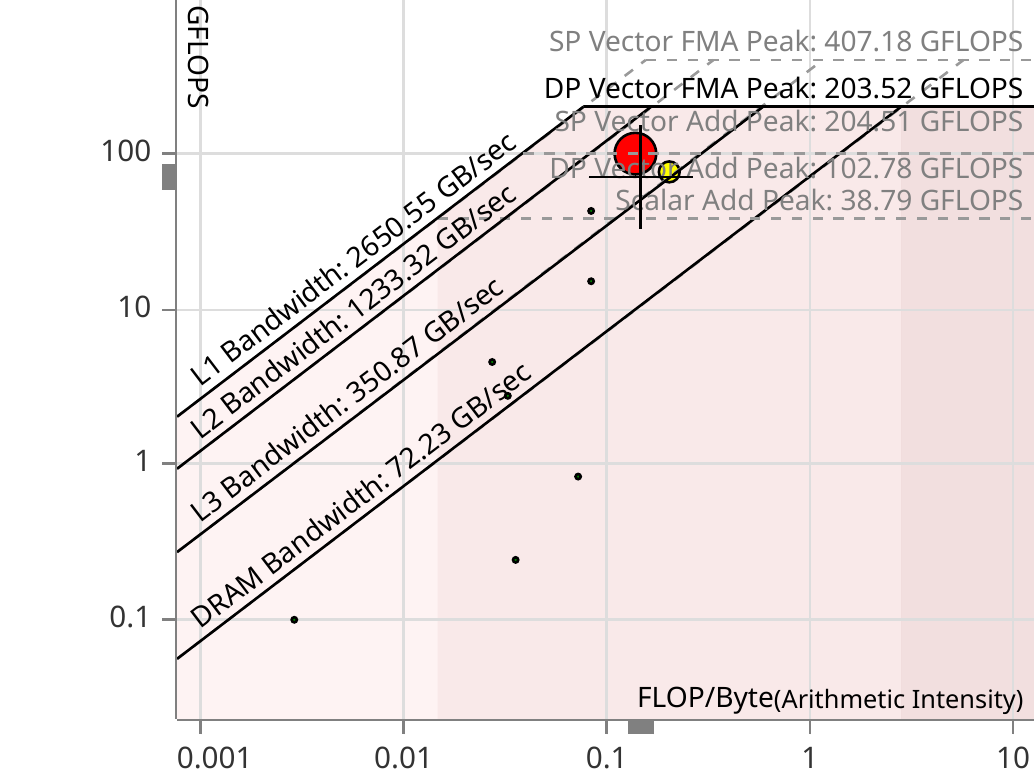}
    \caption{Roofline performance model of the molecular-orbital computation kernels. 
    The red and yellow markers correspond to the dominant computational loops, and the cross shows the average performance.}
    \label{fig:mo_roofline}
\end{figure}

The performance of the kernels involved in the computation of molecular orbitals was evaluated on an Intel Core i7-13850HX processor. The benchmark system is identical to that used for the atomic-orbital (\ac{AO}) benchmarks, namely the Cu(II)–peptide complex described with the cc-pVDZ basis set. This system contains 436 electrons and 1053 molecular orbitals, each expanded over 1114 atomic orbitals.
Figure~\ref{fig:mo_roofline} presents the roofline performance model for the MO computation kernels, comparing the measured throughput with the theoretical bandwidth and compute ceilings of the processor. The dominant computational loops, shown in red and yellow, reach approximately 50\% of the double-precision peak performance, placing it in the compute-bound regime. This indicates that the kernel efficiently exploits vectorization and benefits from good data locality.

\subsection{Molecular orbital cusp correction}

In all-electron calculations, the local energy diverges unless the wave function satisfies the electron–nucleus cusp condition~\cite{kato_1957}. 
In Gaussian-based wave functions, this condition is often imposed through the Jastrow factor. 
An alternative approach, particularly useful in Jastrow-free QMC calculations, is to impose the cusp condition directly on the \acp{MO}, as implemented for example in CASINO~\cite{ma_2005}.

In QMCkl, this is achieved by modifying each \ac{MO} in the vicinity of a nucleus. 
When the electron–nucleus distance becomes smaller than a prescribed radius $r^{\text{cusp}}_{A}$, the \ac{MO} is replaced locally by a function that enforces the correct cusp while ensuring continuity of the orbital and its derivatives. 
For a nucleus $A$, the modified orbital reads
\[
\phi^{\text{cusp}}_{i}(\mathbf{r}) 
= \phi_i(\mathbf{r}) - \phi^{s_A}_{i}(\mathbf{r}) 
+ \sum_{k=0}^{3} f_k\, \|\mathbf{r}-\mathbf{R}_A\|^k ,
\]
where $\phi^{s_A}_{i}$ denotes the contribution to \ac{MO} $i$ from the $s$-type \acp{AO} centered on $A$. 
The coefficients $f_k$ are determined to enforce the continuity of both the values and gradients at $r^{\text{cusp}}_{A}$, while satisfying the exact electron–nucleus cusp condition.

\subsection{Inverse matrices}

Matrix inversion is a fundamental computational kernel in \ac{QMC} simulations. It plays 
a central role in evaluating wave function ratios within the Metropolis algorithm and is 
also essential for computing the gradients and Laplacian of Slater determinants.

The inverse of an $n \times n$ matrix $\mathbf{A}$ is defined as
\begin{equation}
\mathbf{A}^{-1} = \frac{1}{\det(\mathbf{A})} \text{adj}(\mathbf{A}),
\end{equation}
where $\text{adj}(\mathbf{A})$ denotes the adjugate matrix -- the transpose of the cofactor 
matrix of $\mathbf{A}$.

Naively, computing the inverse requires evaluating the determinant, which scales as 
$\mathcal{O}(n!)$, making it infeasible for large $n$. However, efficient algorithms 
exist: by first performing an LU decomposition of $\mathbf{A}$, the inverse can be 
computed in $\mathcal{O}(n^3)$ time by solving $\mathbf{A}^{-1} \mathbf{L} = 
\mathbf{U}^{-1}$. This is typically implemented using LAPACK routines \texttt{getrf} 
(for LU factorization) and \texttt{getri} (for matrix inversion).

Two important observations arise from this approach:
\begin{enumerate}
    \item Once the LU decomposition is computed via \texttt{getrf}, the determinant can 
be evaluated in $\mathcal{O}(n)$ time after factorization. This is because $\mathbf{L}$ is lower triangular 
with unit diagonal entries, and $\mathbf{U}$ is upper triangular, so $\det(\mathbf{A}) = 
\prod_{i=1}^n U_{ii}$.
    
    \item if $\mathbf{r}$ denotes all electron coordinates, the gradient of a Slater determinant $D(\mathbf{r})$ with respect to electron 
position $\mathbf{r}_i$ is commonly expressed as
    \begin{equation}
    \nabla_i D(\mathbf{r}) = D(\mathbf{r}) \sum_{j=1}^n \nabla \phi_j(\mathbf{r}_i) \, 
A^{-1}_{ji}(\mathbf{r}).
    \end{equation}
    Using the adjugate matrix, this can be rewritten as
    \begin{equation}
    \nabla_i D(\mathbf{r}) = \sum_{j=1}^n \nabla \phi_j(\mathbf{r}_i) \, 
\left(\text{adj}(\mathbf{A})\right)_{ji}(\mathbf{r}),
    \end{equation}
    which avoids the explicit multiplication by the determinant. Therefore, it is 
advantageous to return both the determinant and the adjugate of a matrix simultaneously.
\end{enumerate}

The computation of determinants and inverses of \emph{small} matrices (typically $n < 
5$) is also a critical component in many QMC algorithms. For instance, large \ac{CI} 
expansions with a small number of electrons~\cite{caffarel_2016,scemama_2020} or delayed update 
strategies for larger systems~\cite{mcdaniel_2017} rely heavily on efficient 
small-matrix operations. Notably, for $n < 5$, the factorial scaling $\mathcal{O}(n!)$ 
of the naive algorithm becomes more efficient than the $\mathcal{O}(n^3)$ complexity of 
LU-based methods.

\begin{figure}
    \centering
\includegraphics[width=\columnwidth]{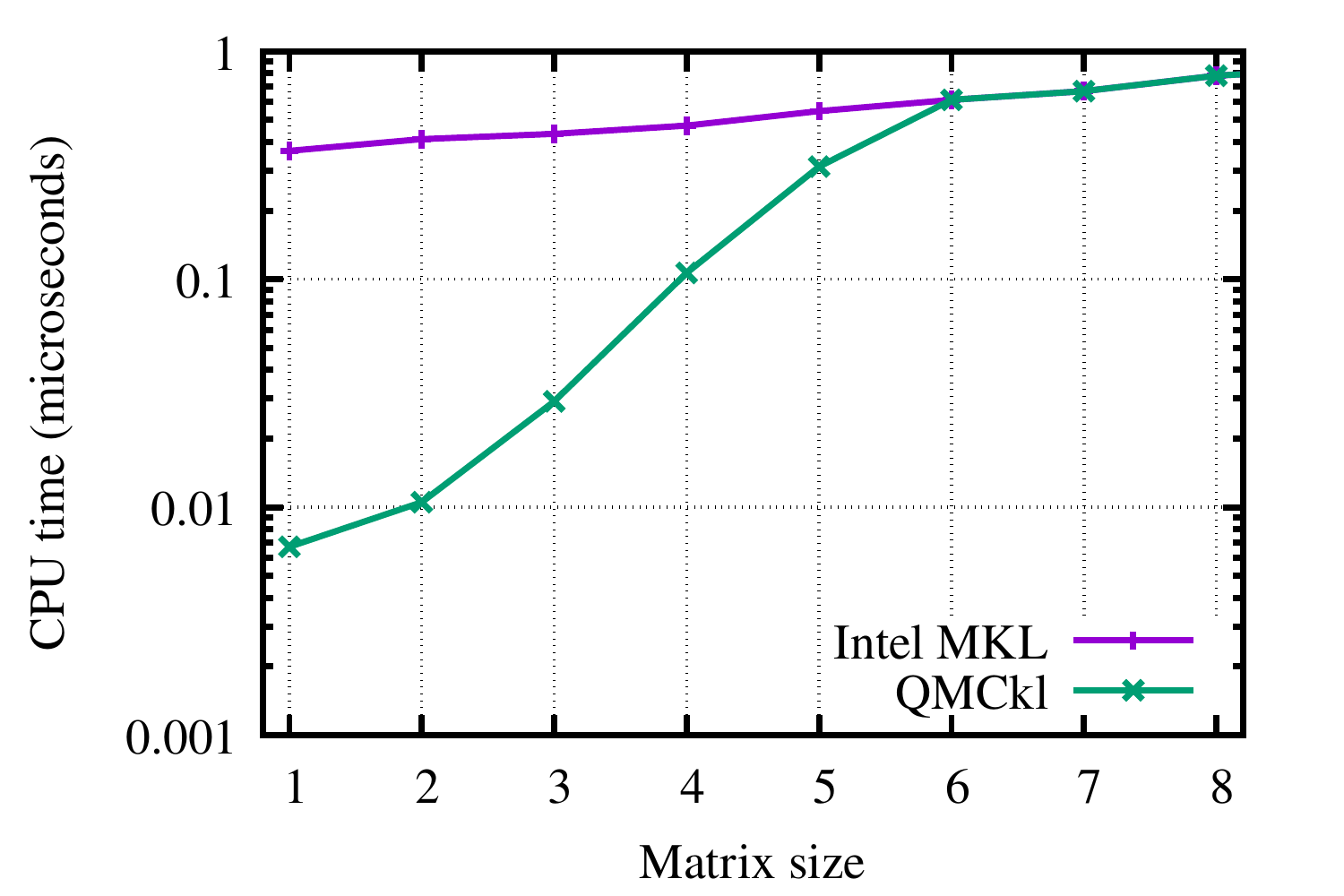}
    \caption{CPU time, in microseconds, needed to compute the adjugate of an $n \times n$ matrix using Intel MKL (\texttt{dgetrf/dgetri}), or QMCkl's specialized functions.}
    \label{fig:det}
\end{figure}

To optimize performance, we have implemented specialized routines in the QMCkl 
library for computing the determinant and adjugate of small matrices ($n < 5$). The 
library automatically switches to these optimized functions when appropriate. 
Performance benchmarks were conducted on a single Intel Xeon Gold 6130 @ 2.10 GHz core. 
QMCkl was compiled using Intel OneAPI 2024.0.1, with \texttt{icx} for C and 
\texttt{ifort} for Fortran, and the Intel MKL library was used for BLAS and LAPACK 
routines (\texttt{getrf}/\texttt{getri}). The results are presented in 
Fig.~\ref{fig:det}, demonstrating the efficiency gain of our specialized implementations 
for small matrices.

\subsection{Jastrow Factor}
To capture dynamical electron correlation, the Slater determinantal part is usually multiplied by a Jastrow factor. The Jastrow factor implemented in QMCkl has the same functional form as in CHAMP~\cite{guclu_2005}, consisting of two-body (electron-electron and electron-nucleus) and three-body (electron-electron-nucleus) components,
\begin{equation}
    J(\mathbf{r}, \mathbf{R}) =  J_\text{en}(\mathbf{r},\mathbf{R}) + J_\text{ee}(\mathbf{r}) + J_\text{een}(\mathbf{r},\mathbf{R}),
\end{equation}
where $\mathbf{r}$ and $\mathbf{R}$ are the electron and nucleus positions, respectively. 

The electron-electron and electron-nucleus contributions are given by
\begin{equation}
    J_\text{en}(\mathbf{r},\mathbf{R}) = \sum_{i=1}^{N_\text{elec}} \sum_{\alpha=1}^{N_\text{nucl}}\left[\frac{a_0 \widetilde{R}_{i\alpha}}{1+a_1 \widetilde{R}_{i\alpha}} + \sum_{k=2}^{N^a_\text{ord}}a_k(\widetilde{R}_{i\alpha})^k\right]
\end{equation}
and 
\begin{equation}
    J_\text{ee}(\mathbf{r}) = \sum_{i=1}^{N_\text{elec}} 
\sum_{j=1}^{i-1}\left[\frac{b_0 \widetilde{r}_{ij}}{1+b_1 \widetilde{r}_{ij}} + \sum_{k=2}^{N^b_\text{ord}}b_k(\widetilde{r}_{ij})^k \right],
\end{equation}
where $N^a_\text{ord}$ and $N^b_\text{ord}$ denote the maximum polynomial order, and the coefficients $a_k$ and $b_k$ are variational parameters. The rescaled distances $\widetilde{r}_{ij}$ and $\widetilde{R}_{i\alpha}$ are defined as
\begin{equation}
    \widetilde{r}_{ij} = \frac{1-e^{-\kappa r_{ij}}}{\kappa} \quad \text{and} \quad  \widetilde{R}_{i\alpha} = \frac{1-e^{-\kappa R_{i\alpha}}}{\kappa},
\end{equation}
where $\kappa$ is a rescaling factor. To ensure that the Jastrow factor vanishes at large separation, the asymptotic contribution $J^\infty_{en}$ and $J^\infty_{ee}$ are subtracted. These are obtained by substituting the asymptotic value of the rescaled distances: $\widetilde{r}^\infty_{ij} = 1/{\kappa}$ and $\widetilde{R}^\infty_{i\alpha} = 1/{\kappa}$.

The three-body Jastrow is defined as
\begin{equation}
\begin{split}
  &J_\text{een}(\mathbf{r},\mathbf{R}) = \sum_{\alpha=1}^{N_\text{nucl}} \sum_{i=1}^{N_\text{elec}} \sum_{j=1}^{i-1}
\sum_{p=2}^{N_\text{ord}^c} \sum_{k=0}^{p-1}
\sum_{l=0}^{p-k-2\delta_{k,0}} c_{lkp\alpha}\\
\times&
\left( {\widetilde{r}}_{ij} \right)^k
\left[ \left( {\widetilde{R}}_{i\alpha} \right)^l + \left( {\widetilde{R}}_{j\alpha} \right)^l \right]
\left( {\widetilde{R}}_{i\,\alpha} \, {\widetilde{R}}_{j\alpha} \right)^{(p-k-l)/2},
\label{J_orig}
\end{split}
\end{equation}
and, for this contribution, the rescaling is modified to enforce a shorter-ranged behavior:
\begin{equation}
    \widetilde{r}_{ij} = e^{-\kappa r_{ij}} \quad \text{and} \quad  \widetilde{R}_{i\alpha} = e^{-\kappa R_{i\alpha}}.
\end{equation}
The straightforward implementation of $J_\text{een}$ as written in Eq.~\eqref{J_orig} scales as $\order{N_\text{elec}^2 \times N_\text{nucl} \times  N_\text{ord}^3}$.

To improve computational efficiency, in QMCkl, the three-body contribution is reformulated in terms of an intermediate matrix $P$ as
\begin{equation}
\begin{split}
   J_\text{een} = &\sum_{p=2}^{N^c_{\text{ord}}}
    \sum_{k=0}^{p-1}
     \sum_{l=0}^{p-k-2\delta_{k,0}}
      \sum_{\alpha=1}^{N_{\text{nucl}}} c_{lkp\alpha}\\
       &\ \times\sum_{i=1}^{N_{\text{elec}}}
       (\widetilde{R}_{i\alpha})^{(p-k+l)/2}
             P_{i\alpha}^{k, (p-k-l)/2},
    \label{J_new}
\end{split}
\end{equation}
where
\begin{equation}
    P_{i\alpha}^{k,m} = \sum_{j=1}^{N_{\text{elec}}}(\widetilde{r}_{ij})^k(\widetilde{R}_{j\alpha})^m.
\end{equation}
A complete derivation of this expression is provided in Sec.~A of the SI. The computational complexity of evaluating $P_{i\alpha}^{km}$ is $\order{N_\text{elec}^2 \times N_\text{nucl} \times N_\text{ord}^2}$, dominating for large $N_\text{elec}$ the cost of evaluating the sums in Eq.~\eqref{J_new}, which scales as $\order{N_\text{elec} \times N_\text{nucl} \times  N_\text{ord}^3}$. This represents a speedup of $N_\text{ord}$ compared to the naive evaluation of the Jastrow factor. Furthermore, the computation of $P_{i\alpha}^{km}$ reduces to a matrix multiplication, allowing the use of highly optimized BLAS matrix-multiplication routines such as \texttt{gemm}. 
Since the Jastrow factor is a short-range operator, its associated matrix representations are potentially highly sparse, especially in large-scale simulations involving extended systems. To exploit this structural property, we have implemented optimized sparse matrix multiplication routines within QMCkl. These routines dynamically select the most efficient algorithm, either sparse or dense, based on the sparsity density of the input arrays at runtime.

\begin{figure}
    \centering
    \includegraphics[width=0.9\linewidth]{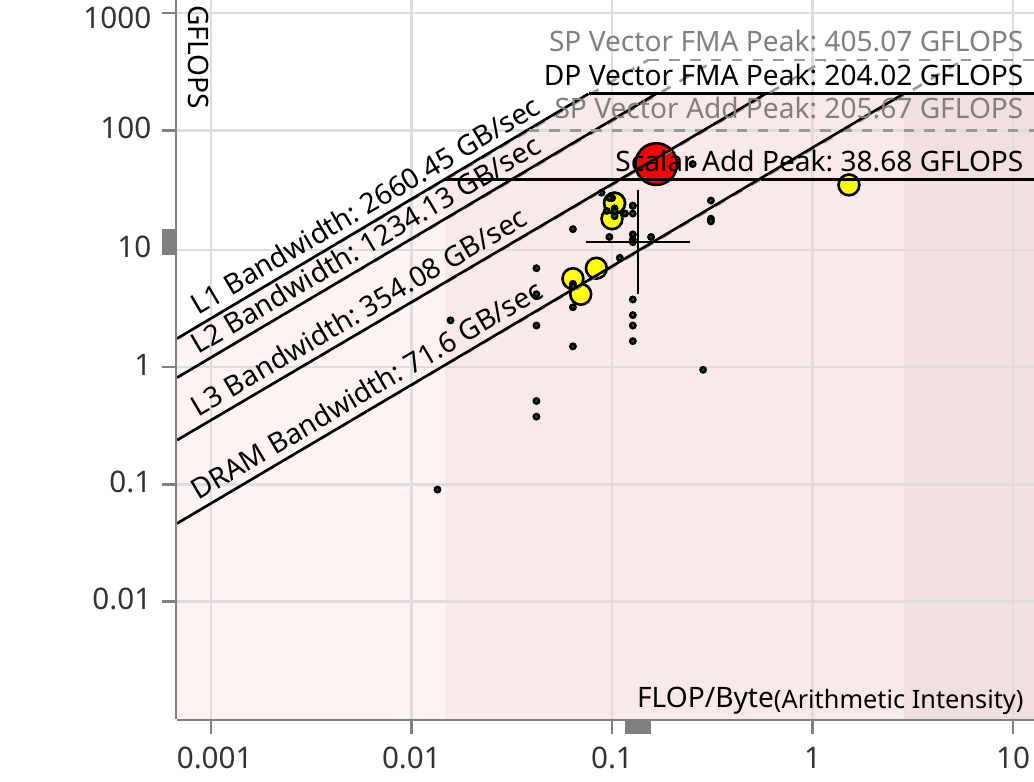}
    \caption{Roofline performance model of the full Jastrow-factor computation kernels, up to three-body terms.
    The red marker denotes the dominant sparse-matrix-multiplication loop in the three-body Jastrow, and yellow markers represent secondary loops. The cross shows the average performance.}
    \label{fig:jastrow_roofline}
\end{figure}

The performance of the kernels responsible for the evaluation of Jastrow factor was measured on the same Intel Core i7-13850HX processor used for the previous benchmarks. The Jastrow factor contains one- and two-electron terms, with powers up to the fifth order in three-body components. The test system is again the Cu(II)–peptide  complex in the cc-pVDZ basis.
Figure \ref{fig:jastrow_roofline} presents the corresponding roofline model, which compares the achieved performance of the Jastrow kernels with the theoretical compute and bandwidth ceilings of the processor.
Most computational loops (yellow markers) lie in the region bounded by the L2 and L3 bandwidth rooflines, indicating a mixed regime where both computation and memory traffic limit performance. The main kernel, shown in red, is the dense-sparse matrix multiplication kernel, which achieves a sustained throughput close to 30\% of the peak performance. Overall, the Jastrow evaluation kernels show balanced behavior, efficiently exploiting both the compute units and the cache hierarchy.

%In \textsc{QMCkl}, the GEMM (GEneral Matrix Multiply) subroutine is used which is one of the fast BLAS (Basic Linear Algebra Subroutines) routines.

Since the $J_\text{ee}$ and $J_\text{een}$ factors depend on all the electron-electron distances, special care is required in calculations where only the coordinates of a single electron differ from the previous electron configuration (e.g. in single-electron Monte Carlo moves and the treatment of effective core potentials). In QMCkl, only the terms of $J_\text{ee}$ and $J_\text{een}$ affected by the move are recomputed. For the three-body Jastrow factor, this is achieved by evaluating the Jastrow before and after the single-electron move,
\begin{equation}
\begin{split}
   \delta J_\text{een} = &\sum_{p=2}^{N^c_{\text{ord}}}
    \sum_{k=0}^{p-1}
     \sum_{l=0}^{p-k-2\delta_{k,0}}
      \sum_{\alpha=1}^{N_{\text{nucl}}} c_{lkp\alpha}\\
       &\ \times \left((\widetilde{R}_{k\alpha}^{\text{new}})^{(p-k+l)/2}\sum_{i=1}^{N_{\text{elec}}} (\widetilde{r}_{ik}^{\text{new}})^k (\widetilde{R}_{i\alpha}^{\text{old}})^{(p-k-l)/2} \right. \\
       &\ + (\widetilde{R}_{k\alpha}^{\text{new}})^{(p-k-l)/2}\sum_{i=1}^{N_{\text{elec}}}(\widetilde{r}_{ik}^{\text{new}})^k (\widetilde{R}_{i\alpha}^{\text{old}})^{(p-k+l)/2} \\
        &\ - (\widetilde{R}_{k\alpha}^{\text{old}})^{(p-k+l)/2}\sum_{i=1}^{N_{\text{elec}}}(\widetilde{r}_{ik}^{\text{old}})^k (\widetilde{R}_{i\alpha}^{\text{old}})^{(p-k-l)/2} \\
        &\ \left. - (\widetilde{R}_{k\alpha}^{\text{old}})^{(p-k-l)/2}\sum_{i=1}^{N_{\text{elec}}}(\widetilde{r}_{ik}^{\text{old}})^k (\widetilde{R}_{i\alpha}^{\text{old}})^{(p-k+l)/2} \right),\\
\end{split}
\end{equation}
where $k$ is the index of the moved electron. Therefore, in a single-electron move, the intermediate matrix $P$ is not explicitly updated because computing its change is highly memory intensive: rank-one update have a low arithmetic intensity and are therefore memory-bound. Under these conditions, we have found that exploiting sparsity directly in \eqref{J_orig} is more efficient.

As with the orbital routines, QMCkl provides not only the values of the Jastrow factor but also its gradients and Laplacians with respect to the electron coordinates, both for the full set of coordinates and for single-electron updates. In addition, derivatives with respect to the Jastrow parameters are implemented, enabling wave function optimization. 

\subsection{Interatomic Forces}

Finally, QMCkl provides all the necessary kernels for the calculation of interatomic forces, which are essential for structural relaxation and molecular dynamics within QMC. The low-variance expression of the estimator of the force, valid both in variational Monte Carlo (VMC) and diffusion Monte Carlo (DMC), is given by
\begin{equation}
    F = -\partial_{\alpha} E = -\expval{\partial_{\alpha} E_L + (E_L - E) \partial_{\alpha}P}_P,
\end{equation}
where $\partial_\alpha$ denotes the derivative with respect to the nuclear coordinates, and $E_L=H\Psi/\Psi$ is the local energy. In VMC, the probability distribution $P$  is simply proportional to $|\Psi(\mathbf{r})|^2$ (the corresponding DMC formulation can be found in Ref.~\citenum{Moroni2014}). In general, force evaluation requires the derivatives with respect to nuclear coordinates of the wave function, the gradient of the wave function with respect to the electron coordinates, and of the local energy. This in turn involves up to third-order derivatives of the wave function. 

\section{Integration in Quantum Chemistry Codes and Performance}\label{sec:integration}
\begin{figure}
    \centering
    \includegraphics[width=1\linewidth]{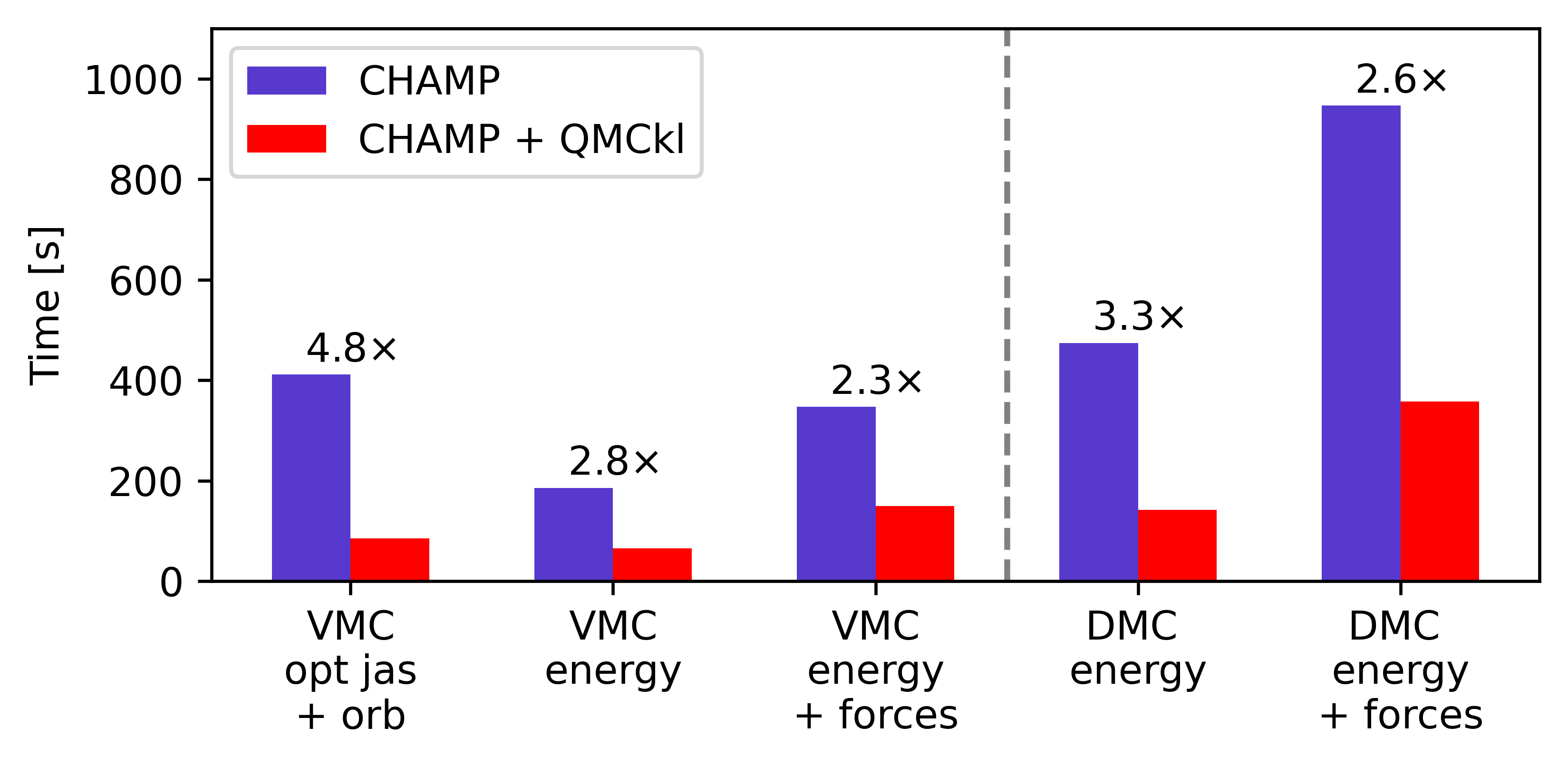}
    \caption{Computational speedup in CHAMP using QMCkl for various common usage modes, demonstrated on the molecule aspirin with a three-body Jastrow factor. All benchmarks were performed on 16 AMD EPYC 9654 CPUs @ 2.40 GHz (1536 cores). QMCkl and CHAMP are compiled using Intel OneAPI 2023.1.0, with icx for C and ifort for Fortran, and the Intel MKL library is used for BLAS and LAPACK routines.}
    \label{fig:champ_speedup}
\end{figure}

The QMCkl library has been integrated into the QMC codes QMC=Chem and CHAMP, and in the wavefunction code Quantum Package. Each program exploits different kernels, reflecting the modular design of QMCkl, and benefits from improved performance, accuracy, and interoperability through the common TREXIO format.

\subsection{QMC=Chem}

In QMC=Chem, QMCkl kernels are used for the evaluation of atomic and molecular orbitals, cusp-corrected molecular orbitals, and the Jastrow factor. While QMC=Chem originally relied on single-precision routines, QMCkl provides also efficient double-precision implementations, increasing accuracy and cross-code validation. The adoption of the CHAMP-compatible Jastrow factor allows seamless transfer of wave functions between CHAMP and QMC=Chem, for example for optimizing the Jastrow in CHAMP and then performing large-scale multi-determinant calculations in QMC=Chem. 

\begin{table}
\caption{\label{tab:qmcchem} Single-core performance benchmark of QMC=Chem with and without QMCkl across different compilers and architectures. }
\begin{tabular}{lccrr}
\hline
CPU & QMC=Chem & QMCkl & Time& Speedup\\
    & compiler & compiler & (ms) & \\
\hline
Intel Core i7 & Intel  &   no QMCkl          & 26.69 & --- \\
              & Intel  & GNU   & 20.58 & $1.30\times$ \\
              & Intel  & Intel & 19.86 & $1.34\times$ \\
\hline
Intel Core i7 & GNU   &   no QMCkl          & 30.53 & --- \\
              & GNU   & GNU   & 20.51 & $1.49\times$ \\
              & GNU   & Intel & 19.56 & $1.56\times$ \\
\hline
 ARM          & GNU & no QMCkl & 41.71 & --- \\
Neoverse V1   & GNU & GNU      & 21.67 & $1.92\times$ \\
              & GNU & ARM      & 22.42 & $1.86\times$ \\
\hline
\end{tabular}
\end{table}

We report here single-core performance benchmarks of QMC=Chem, both with and without integration of the QMCkl library, across different compilers (Intel \texttt{ifx} Fortran compiler version 2023.2.4, GNU \texttt{gfortran} and \texttt{gcc} version 13.3.0, and ARM \texttt{armclang} C compiler version 24.10.1) and hardware architectures (x86: Intel Core i7-13850HX, and ARM Neoverse V1: Ampère Computing Altra Q80-30). The benchmark system is a Hartree-Fock determinant of the C$_{60}$ molecule using the cc-pVQZ basis set and \ac{BFD} \acp{ECP}\cite{Burkatzki2007,BFD_H2013}. This system comprises 4140 atomic orbitals, 120 molecular orbitals, and 240 electrons, and is used as a benchmark for the computation of the local energy at a single Monte Carlo configuration. 

The results in Table~\ref{tab:qmcchem} highlight a critical limitation of QMC=Chem: performance is sensitive to the choice of compiler. When compiled with Intel’s \texttt{ifx} Fortran compiler, QMC=Chem achieves optimal performance on x86 architectures due to aggressive vectorization and memory alignment optimizations using vendor-specific directives. However, this performance advantage does not extend to other compilers nor to non-x86 platforms like ARM, where performance degrades.

In contrast, when QMCkl is used, regardless of the QMC=Chem compiler, the performance is consistently superior and highly portable. QMCkl’s optimized kernels enable near-optimal performance across all tested configurations. On the Intel Core i7, replacing the native Fortran routines with QMCkl compiled with \texttt{gcc} or \texttt{icx} yields speedups of $1.30\times$ to $1.56\times$. On the ARM platform, the speedup is even more dramatic: up to $1.92\times$ with \texttt{gcc} and $1.86\times$ with \texttt{armclang}, demonstrating that QMCkl effectively bridges the performance gap between architectures.

\subsection{CHAMP}

In CHAMP, QMCkl is used for the evaluation and optimization of molecular orbitals and the Jastrow factor. Both operations support single-electron moves as well as the treatment of non-local effective core potentials (ECPs). Furthermore, calculation of interatomic forces is also possible through QMCkl. 
In Fig.~\ref{fig:champ_speedup}, we show the speedup of the code with respect to the original implementation for calculations on the aspirin molecule (21 atoms, 68 electrons) using a three-body Jastrow factor. We employ the \ac{BFD} \acp{ECP} and the corresponding cc-pVTZ basis set. In the DMC calculations, the non-local ECPs were treated with the size-consistent T-move algorithm\cite{Casula2006} with an additional accept/reject step\cite{Anderson2021}. 

In VMC, QMCkl accelerates energy-only calculations by $2.8\times$ and combined energy–force evaluations by $2.3\times$. The largest improvement occurs during wave function optimization, where the computation of the wave function derivatives and of the Hamiltonian acting on them achieves a $4.8\times$ speedup. In DMC, energy and combined energy–force evaluations are accelerated by $3.3\times$ and $2.6\times$, respectively.

\begin{figure*}[t!]
    \centering
    \begin{subfigure}[t]{0.5\textwidth}
        \centering
        \includegraphics[width=\textwidth]{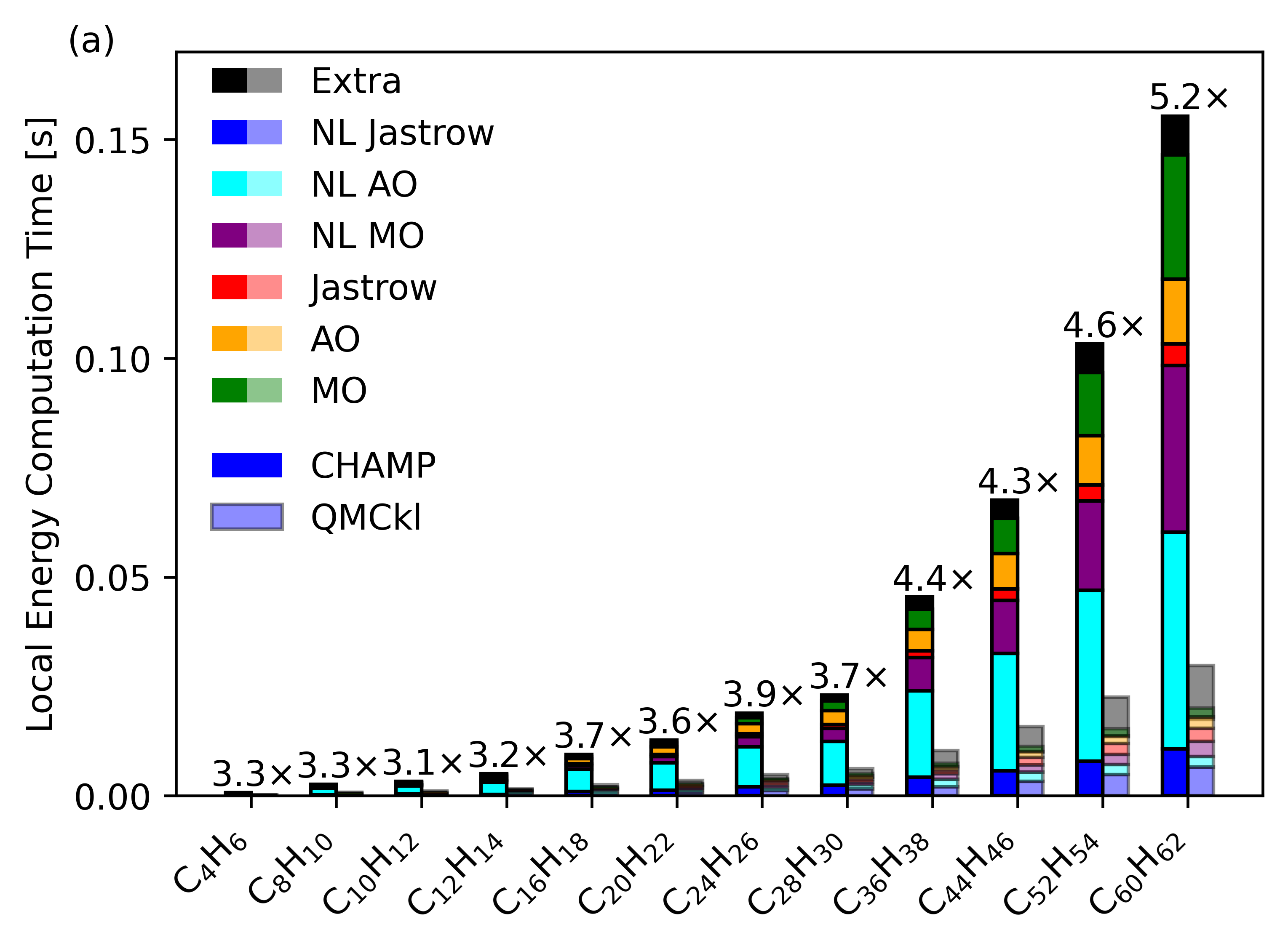}
    \end{subfigure}%
    ~ 
    \begin{subfigure}[t]{0.5\textwidth}
        \centering
        \includegraphics[width=\textwidth]{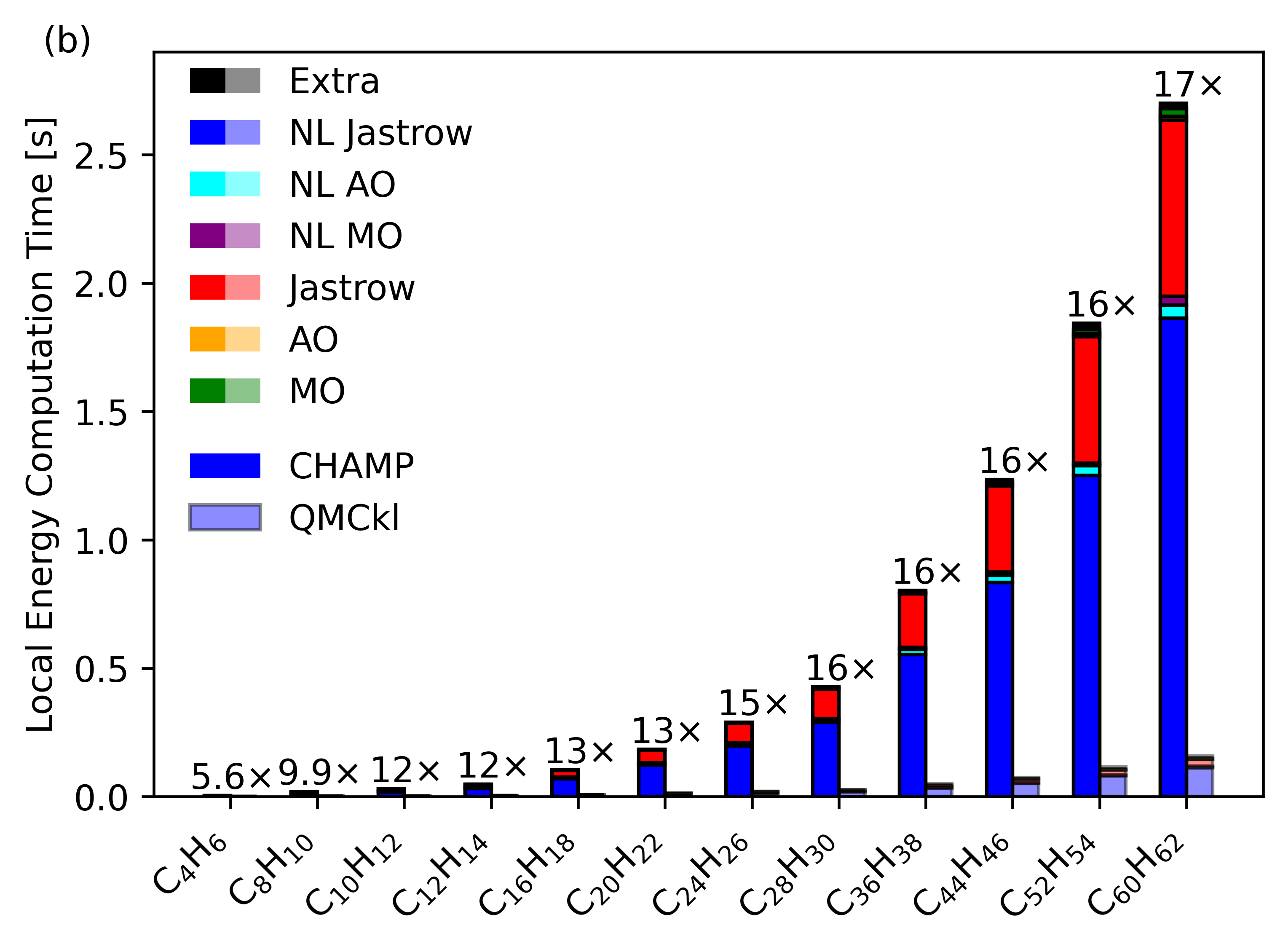}
    \end{subfigure}
    \caption{Breakdown of computational time (s) for evaluating the local energy in CHAMP with (shaded) and without (solid) QMCkl for polyenes of increasing size. The achieved speedup is shown above the bars. Results are given for (a) a two-body Jastrow factor with $\kappa = 0.6$ a.u. and (b) an additional short-range three-body Jastrow with $\kappa = 2.0$ a.u. ``NL'' denotes the nonlocal \ac{ECP} integration, and ``extra'' the remaining parts (e.g., Slater determinant). Benchmarks were performed on a single Intel Xeon Gold 6326 core @ 2.90 GHz. QMCkl and CHAMP are compiled using Intel OneAPI 2021.4.0, with icx for C and ifort for Fortran, and the Intel MKL library is used for BLAS and LAPACK routines.
    %Breakdown of the computational time (s) for the evaluation of the local energy in CHAMP with (shaded bars) and without (full bars) QMCkl for increasingly larger polyenes. The speedup achieved using QMCkl is reported above the bars. Results are shown for (a) a two-body Jastrow factor with a rescale factor $\kappa = 0.6$ a.u.\ and (b) a short-range three-body Jastrow factor with a large rescale factor $\kappa = 2.0$ a.u. ``NL'' denotes the integration of the nonlocal ECP requiring many single-electron moves, and ``extra'' the remaining parts of the calculation such as the evaluation of the Slater determinant. All benchmarks are performed on a single core of an Intel Xeon Gold 6326 CPU (2.90 GHz).
    }
    \label{fig:benchmarks}
\end{figure*}

In Fig.~\ref{fig:benchmarks}, we show the breakdown of the computational time for the local energy evaluation in CHAMP with and without QMCkl for increasingly larger polyenes. All calculations employ \ac{BFD} \acp{ECP} with a cc-pVDZ basis. In the two-body Jastrow case [Fig.~\ref{fig:benchmarks}(a)], most of the time in CHAMP (without QMCkl) is spent on the evaluation of atomic and molecular orbitals. Enabling QMCkl leads to a substantial performance improvement (up to about $5.2\times$ for the largest C$_{60}$H$_{62}$), mainly due to the significant acceleration in the numerical integration of the nonlocal \ac{ECP}. When a three-body Jastrow factor is employed [Fig.~\ref{fig:benchmarks}(b)], the computational cost of the local energy becomes dominated by the Jastrow evaluation. In this regime, QMCkl provides an even greater speedup for larger polyenes (up to $17\times$) due to its optimized Jastrow implementation.

\subsection{Quantum Package}

One of the key applications of QMCkl lies in the transcorrelated selected configuration interaction (TC-CIPSI) methods developed within Quantum Package.~\cite{ammar_2022} In this framework, the electronic Hamiltonian is transformed using a Jastrow factor, improving the short-range treatment of electron correlation. However, this transformation also leads to a substantial increase in the computational cost associated with evaluating the modified matrix elements.
Quantum Package addresses this bottleneck by leveraging QMCkl in the numerical evaluation of integrals involving Jastrow factors.
As a result, the CHAMP code can be employed to optimize the parameters of the Jastrow factor used in transcorrelated calculations performed with Quantum Package. This interoperability highlights the modular design philosophy of QMCkl: its kernels can be reused across distinct but complementary software ecosystems.

\section{Usage Examples} \label{sec:examples}

\subsection{Example in Python}

The following Python example illustrates how to create a QMCkl context, read wave function data from a TREXIO file, and evaluate molecular orbitals at a set of points. Examples in Fortran and C are available in Sec.~B of the SI.  

First, a context is created and populated with the wave function information stored in the TREXIO file. Next, the number of MOs is retrieved to determine the appropriate array size for their evaluation. The positions of the points at which the MOs will be evaluated are then supplied to QMCkl. These points typically correspond to electron coordinates or grid points, and are generated by the calling quantum chemistry code. Once the points are set, the MOs can be evaluated at these positions. The results are stored in arrays accessible to the calling program. Finally, the context is destroyed to free all allocated memory.
\begin{minted}[bgcolor=white,fontsize=\footnotesize, frame=lines]{Python} 
import qmckl
import numpy as np

# Create a QMCkl context
context = qmckl.context_create()
  
# Read the TREXIO file into the context
trexio_filename = <supplied by package>
qmckl.trexio_read(context, trexio_filename)

# Retrieve the number of MOs
mo_num = qmckl.get_mo_basis_mo_num(context)

point_num = <supplied by package>
point = <supplied by package>

# Set electron positions
qmckl.set_point(context, 'N', point_num, 
                np.reshape(point, (point_num*3)))

# get the MO values
mo_value = qmckl.get_mo_basis_mo_value(context, 
                              point_num*mo_num)

qmckl.context_destroy(context)
\end{minted}
Additional examples, installation instructions, and code documentation are available at the QMCkl GitHub\cite{qmckl_github}.

\subsection{Ensuring consistency of the Jastrow factor across codes}
\label{sec:consistency} 

In this section, we demonstrate the practical benefits of using QMCkl in a multi-code workflow by optimizing a Jastrow factor for the nitrogen molecule using VMC in CHAMP, and subsequently employing the same optimized Jastrow factor in transcorrelated (TC) calculations within Quantum Package.  

A key challenge in such workflows arises from the non-variational nature of the transcorrelated energy: unlike the variational energy, it cannot be computed deterministically, making it difficult to verify that the Jastrow factor used in one code is identical to the one used in another. This consistency can be guaranteed if both codes rely on the same underlying computational implementation. In our case, this is achieved through the shared use of QMCkl for evaluating the Jastrow factor. Specifically, Quantum Package computes the required integrals involving the Jastrow factor numerically using the routines provided by QMCkl, ensuring that the functional form and evaluation procedure are identical across both codes. 

To rigorously validate this consistency, we propose a cross-checking protocol, taking advantage of the fact that QMC=Chem can sample both the variational and transcorrelated energies:
\begin{enumerate}
    \item Optimize the Jastrow factor in CHAMP.
    \item Verify that the variational energy computed in QMC=Chem (which also uses QMCkl) is in agreement with the one obtained in CHAMP.
    \item Sample the transcorrelated energy in QMC=Chem, using the same Jastrow factor.
    \item Compute the transcorrelated energy in Quantum Package, using the same Jastrow factor.
    \item Compare the transcorrelated energy obtained in Quantum Package with the one computed in QMC=Chem.
\end{enumerate} 

The results, summarized in Table~\ref{tab:qmc_tc}, confirm that the transcorrelated energy computed in Quantum Package is fully consistent with the one obtained in CHAMP, validating that the Jastrow factor used in both codes is identical. This demonstrates the power of QMCkl as a shared computational foundation for reproducible and interoperable QMC simulations. 

\begin{table}[htbp]
\centering
\caption{Comparison of variational and transcorrelated energies (a.u.) computed with different codes, confirming consistency of the Jastrow factor across workflows.}
\label{tab:qmc_tc}
\begin{tabular}{lll}
\hline
Code            & Quantity                   & Value \\
\hline
CHAMP  & Variational energy         & -109.3039(4) \\
QMC=Chem & Variational energy       & -109.3032(5) \\
QMC=Chem & Transcorrelated energy   & -109.5151(5) \\
Quantum Package & Transcorrelated energy & -109.51538 \\
\hline
\end{tabular}
\end{table} 

The excellent agreement between the energies, within statistical uncertainty, confirms that the Jastrow factor optimized in CHAMP is faithfully reproduced and correctly evaluated in Quantum Package, thanks to the shared use of QMCkl. This workflow exemplifies how the QMCkl library enables reliable cross-code validation.

\subsection{Validation of Basis Set Definitions in TREXIO Files}
\label{sec:trexio_validation} 

Writing a wave function in TREXIO format requires careful adherence to the conventions of the underlying quantum chemistry code, particularly in the definition of basis set normalization factors. Due to the fact that only one- and two-electron integrals are typically needed to compute physical observables, errors in the basis set representation, such as incorrect normalization or inconsistent contraction schemes, are often difficult to detect.

To address this challenge, the \texttt{trexio-tools} package\cite{trexio_tools}
leverages QMCkl to evaluate atomic and molecular orbitals on a spatial grid directly from the data stored in a TREXIO file. Using these numerical evaluations, the overlap matrix is computed via numerical integration. This numerically obtained overlap matrix is then compared with the exact overlap matrix computed by the original quantum chemistry code and stored within the same TREXIO file. A high degree of agreement between the two matrices provides strong evidence that the basis set information was correctly written in the file. 

This validation procedure not only ensures the integrity of wave function data exchange between codes but also serves as a powerful diagnostic tool during the development and testing of new input formats.
%By relying on QMCkl for consistent and accurate orbital evaluation, \texttt{trexio-tools} enables reproducible and trustworthy interoperability across codes.

\subsection{Accelerating Electron Density and Orbital Visualization with QMCkl}

\begin{figure}
    \centering
    \includegraphics[width=\columnwidth]{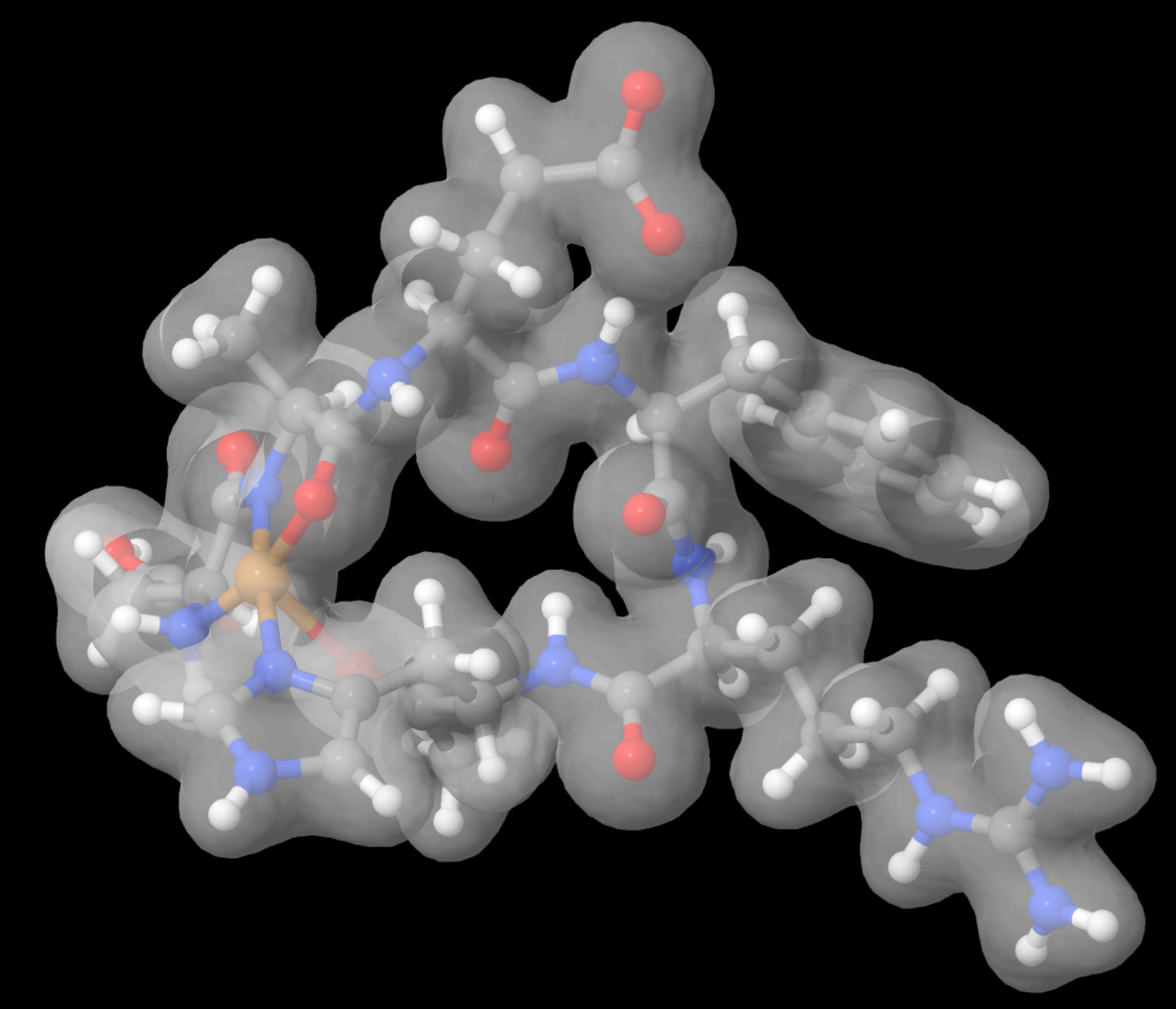}
    \caption{Electron density of the Cu(II)–peptide complex [C$_{34}$H$_{48}$CuN$_{11}$O$_{10}$]$^{2+}$.}
    \label{fig:density}
\end{figure}

The high-performance kernels provided by QMCkl for evaluating molecular orbitals and their derivatives, including gradients and Laplacians, at three-dimensional points are not limited to QMC workflows.
These capabilities can also be used to accelerate the evaluation of orbitals and their spatial derivatives at grid points in density functional theory calculations, as well as in methods that rely on the analysis of the electron density, such as the Electron Localization Function (ELF)~\cite{elf} and the Quantum Theory of Atoms in Molecules (QTAIM)~\cite{qtaim}.

Moreover, QMCkl can offer dramatic speedups in real-time visualization of molecular orbitals and of the electron density. For instance, we demonstrate its utility in computing the electron density of a Kohn-Sham determinant for the Cu(II)–peptide complex complex, expanded in the cc-pVDZ basis set. This system comprises 1114 atomic orbitals and 218 occupied molecular orbitals, resulting in a computationally demanding density evaluation.
We benchmarked the computation of the electron density on a $100 \times 100 \times 100$ Cartesian grid using two approaches:
\begin{itemize}
    \item The standard implementation in Molden~6.9~\cite{schaftenaar_2000} compiled with the GNU compiler,
    \item A lightweight Python script that directly calls QMCkl for orbital evaluation and writes the density into a text file in \texttt{.cube} format.
\end{itemize}

This task, performed in Molden~6.9~\cite{schaftenaar_2000}, required 450 seconds on an Intel Core i7-13850HX CPU. In contrast, our QMCkl-based Python script, handling the full pipeline from reading the TREXIO file, evaluating the density via sum-of-squared molecular orbitals, to writing the output in \texttt{.cube} format, completed in just 10.1 seconds with a single thread, and further reduced to 5.0 seconds using 18 threads. This represents a speedup of over $90\times$ compared to Molden. 

Crucially, for visualization purposes, double precision is not required. By switching to single-precision arithmetic, we achieved even greater performance: the single-threaded version ran in 8.4 seconds, and the 18-threaded version in only 3.7 seconds, yielding a speedup exceeding $120\times$ relative to Molden.

These results highlight the efficiency of QMCkl’s high-performance kernels, especially when combined with multi-threading and reduced-precision arithmetic. The performance gains make it ideal for real-time visualization, large-scale screening, or integration into interactive workflows where speed and responsiveness are critical.

\section{Conclusion} \label{sec:conclusion}
In this work, we presented QMCkl, a modular and portable library providing optimized computational kernels for QMC applications. Its design separates algorithmic clarity from performance optimization, ensuring identical numerical results across platforms and codes. QMCkl delivers significant performance improvements across all stages of a QMC workflow, including the calculation of the energy and interatomic forces, and the optimization of Jastrow-Slater wave functions. 

The modular kernel approach facilitates interoperability across independent QMC codes, such as CHAMP and QMC=Chem, and extends to deterministic quantum chemistry frameworks like Quantum Package, where QMCkl enables the efficient computation of the Jastrow factor in transcorrelated calculations. Furthermore, the orbital routines of QMCkl accelerate tasks such as orbital analysis and visualization, achieving speedups exceeding $100\times$ compared to standard tools. Finally, by combining performance, flexibility, and portability, QMCkl extends the potential applicability of QMC methods and facilitates their integration into a wider range of quantum chemistry frameworks.

\section*{Acknowledgments}

This work received funding from the European Union’s Horizon 2020 research and innovation programme under grant agreement No.~952165 (TREX Center of Excellence in Exascale Computing).
We acknowledge the use of the Dutch national e-infrastructure with the support of the SURF Cooperative (grant No.~NWO-2025.003).
This work was supported by a French government grant managed by the Agence Nationale de la Recherche under the ``Investissements d'avenir'' program (reference ``ANR-21-ESRE-0051''), and was granted access to the MesoNET resources center and the MesoNET Project under the allocation m23005.

\bibliographystyle{aipnum4-2}
\bibliography{qmckl}

\begin{acronym}[TRUST-IT] %width of the longest acronym should be matched here

\acro{AO}{atomic orbital}
\acro{API}{Application Programming Interface}
\acro{BLAS}{Basic Linear Algebra Subprograms}
\acro{BFD}{Burkatzki-Filippi-Dolg}
\acro{CINECA}{Consorzio Interuniversitario}
\acro{CNRS}{Centre National de la Recherche Scientifique}
\acro{CoE}{Center of Excellence}
\acro{CI}{configuration interaction}
\acro{DFT}{Density Functional Theory}
\acro{DMC}{Diffusion Monte Carlo}
\acro{VMC}{Variational Monte Carlo}
\acro{ECP}{Effective Core Potential}
\acro{ERI}{Electron Repulsion Integral}
\acro{EZFIO}{Easy Fortran Input/Output}
\acro{FFI}{Foreign Function Interface}
\acro{FZJ}{Forschungszentrum Jülich GmbH}
\acro{GEMM}{general matrix multiplication}
\acro{GPU}{Graphical Processing Unit}
\acro{HDF5}{Hierarchical Data Format}
\acro{HPC}{High Perfomance Computing}
\acro{HTML}{HyperText Markup Language}
\acro{KTH}{Kungliga Tekniska högskolan}
\acro{LAPACK}{Linear Algebra PACKage}
\acro{MKL}{Math Kernel Library}
\acro{Megware}{Megware computer vertrieb und service GmbH}
\acro{MO}{molecular orbital}
\acro{MPG}{Max Planck Gesellschaft zur forderung der wissenschaften}
\acro{MPI}{Message passing interface}
\acro{PDF}{Portable Document Format}
\acro{QMCkl}{quantum Monte Carlo kernel library}
\acro{QMC}{quantum Monte Carlo}
\acro{SCF}{Self Consistent Field}
\acro{SISSA}{Scuola Internazionale Superiore di Studi Avanzati di trieste}
\acro{STUBA}{Slovenská technická univerzita v Bratislave}
\acro{SAV}{Slovak Academy of Sciences}
\acro{TREX}{Targeting REal chemical accuracy at the eXascale}
\acro{TREXIO}{TREX Input/Output}
\acro{TRUST-IT}{TRUST-IT SRL}
\acro{TUL}{Politechnika Łódzka}
\acro{UNIVIE}{Universität Wien}
\acro{ULP}{unit in the last place}
\acro{UT}{Universiteit Twente}
\acro{UVSQ}{Université de Versailles Saint-Quentin-en-yvelines}
\acro{WP}{Work Package}

\end{acronym}

\end{document}